\documentclass[tradiabstract]{aa}
\usepackage{graphicx}
\usepackage{txfonts}
\newcommand{\HI}{\ion{H}{i}}

\newcommand{\kms}{km sec$^{-1}$}

\newcommand{\ltsima} {$\; \buildrel < \over \sim \;$}
\newcommand{\gtsima} {$\; \buildrel > \over \sim \;$}
\newcommand{\lta} {\lower.5ex\hbox{\ltsima}}
\newcommand{\gta} {\lower.5ex\hbox{\gtsima}}

\newcommand{\OI}{[O{\,\small I}]}

\newcommand{\SII}{[S{\,\small II}]}
\newcommand{\NII}{[N{\,\small II}]}

\newcommand{\HaNII}{[N{\,\small II}+H$\alpha$]}

\begin{document}


\title{Broad H\,{\Large\bf\sffamily I} absorption in the candidate binary black-hole\\ 4C37.11 (B2 0402+379)}
\titlerunning{Broad H\,{\tiny I} absorption in the candidate binary black-hole\\ 4C37.11}
\authorrunning{Morganti et al.}
\author{Raffaella Morganti\inst{1,2}
\and
Bjorn Emonts\inst{3}
\and
Tom Oosterloo\inst{1,2}
}

\institute{Netherlands Institute for Radio Astronomy, 
Postbus 2, 7990 AA, Dwingeloo, The Netherlands
\and
Kapteyn Astronomical Institute, University of Groningen, Postbus 800, 9700 AV Groningen, The Netherlands
\and 
CSIRO Australia Telescope National Facility, PO Box 76, Epping NSW, 1710, Australia
}

\date{Received: Accepted: }

\abstract{We report the discovery of extremely broad 21-cm \HI\ absorption (FWZI
$\sim$1600 \kms) detected with the  Westerbork Synthesis Radio Telescope in the
radio source  4C37.11 (B2 0402+379). This object has been claimed
to host a super-massive binary black hole (Rodriguez et al.\
2006). The main features in the absorption profile are two components, separated
by $\sim$1100 \kms. The  \HI\ absorption in 4C37.11 is unusual because it is the
first case where such broad absorption is found to be {\sl centred} on the
systemic velocity of the host galaxy and not asymmetric and blueshifted as is
seen in all other galaxies with broad \HI\ absorption.    
Given the large width of the absorption, we suggest that a
possible explanation for the extreme properties of the \HI\ absorption is that
it is the kinematic signature of a binary black hole. If this interpretation is
correct, the combined black hole mass derived from the absorption profile is
consistent with that derived from the luminosity of the spheroid.  If the broad
absorption is indeed due to a binary black hole,  this finding confirms the
importance of the gaseous component  in the merging process of supermassive
black holes. }


\maketitle

\section{Introduction}

Galaxy mergers and accretion are  common phenomena, even
in our nearby Universe. Early-type galaxies, and in particular those hosting an
active galactic nucleus (AGN), are believed to form through such mergers or
accretion events. Because most early-type galaxies are known to host a
black-hole (e.g., Ferrarese \& Ford 2005), it is therefore natural to expect
that the two  black holes (BH) from the progenitor galaxies will sink to the
centre of the merger remnant where eventually they will merge. During this
process, a binary black hole system is expected to exist  in the centre of the
newly created galaxy. Recent numerical simulations (see e.g., Mayer et al.\ 2007
and  references therein) have emphasised that this process is  profoundly influenced by
the presence of a gaseous component, because  the orbital evolution of merging 
supermassive BHs is strongly affected by  friction against this gaseous
background. Gas is indeed commonly present in the nuclear regions of AGN galaxies.
This gas often shows complex kinematics  due to the effects of the AGN on the
ISM (e.g., gas outflows; Morganti et al.\ 2005, Holt et al.\ 2008 and refs
therein). Even more extreme kinematics could be expected in the case of binary
black holes, where the gas in the nuclear region may be strongly disturbed and
kinematical signatures of this might be observed.

Identifying binary black holes is not easy because the distance between the two
black holes can be quite small and high spatial resolution is required. Indirect
evidence for spatially unresolved binary black holes can be found, e.g., from
the morphology of the radio emission and radio jets,  or  from semi-periodic
signals in lightcurves (see Komossa 2003 for a review). However, a few cases of
spatially resolved systems are  known. These include NGC~6240 (Komossa et al.\
2003) and 3C~75 (Owen et al.\ 1985). One candidate recently proposed is the
super-massive binary black hole system in the radio galaxy 4C37.11 (B2
0402+379). 
This radio source  is classified as a compact symmetric object (CSO)
although it has some properties that are unusual for a CSO. Interestingly, it is
one of the few CSOs to posses a kiloparsec-scale radio structure.  On VLBA
scales, it has  {\sl two} compact, variable, flat-spectrum active nuclei,
separated by only 7.3 pc. From one of them, two relatively symmetric radio lobes
with a total extent of $\sim$40 pc emanate. Because of this morphology, 4C37.11
has been considered a candidate binary black-hole (Rodriguez et al.\ 2006).

\HI\ absorption was earlier detected in 4C37.11  against one of the radio lobes
by Maness et al.\ (2004). Here we present new \HI\ observations revealing that
the absorption is much broader than found by Maness et al., and showing two
components separated by $\sim$1100 \kms. The characteristics of the broad
absorption in 4C37.11 are different from other, previously detected broad \HI\
absorption  systems, and we discuss whether the absorption in 4C37.11 could
indicate the existence of a binary black hole system in this galaxy.
We also
present optical spectra to  investigate the kinematics  of the ionised gas, and
to confirm the value of the systemic velocity  which is important for the
interpretation of the \HI\ absorption.

\label{sec:radio}
\section{The broad \HI\ absorption}

\begin{figure}
\centering
\includegraphics[width=0.48\textwidth]{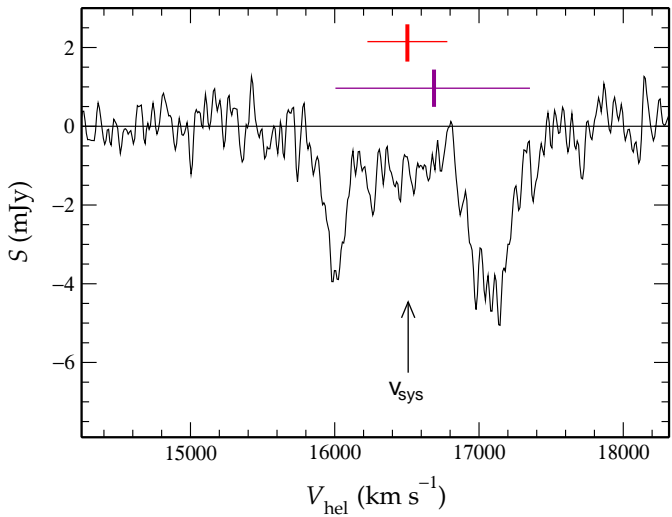}
\caption{\HI\ absorption profile of 4C37.11. The kinematics of the optical \NII\ emission-line are indicated in color; the thick  vertical colored bars indicate the central velocity of the narrow (red) and broad (magenta) component of \NII, while the thin
 horizontal colored bars indicate the FWHM of the narrow and broad component. The systemic velocity ($v_{\rm sys}$) derived from the optical emission-lines is also shown.}
\label{fig:HI}
\end{figure}

\HI\ observations of 4C37.11 were performed using the Westerbork Synthesis Radio Telescope (WSRT) on 1 Nov 2003 (duration 5 hours) and were repeated on 4 June 2005 (duration 11 hours) to confirm the detection of the broad absorption profile.  
We used two bands of 20~MHz bandwidth each, both centred on 1346~MHz and with
512 channels. The data were reduced using the MIRIAD software (Sault et al.\
1995).  The source 4C37.11 is unresolved at the resolution of the WSRT-L band
system (about 12 arcsec). A noise level of 0.39 mJy beam$^{-1}$ was achieved
for a velocity resolution of 36 \kms\ (after Hanning smoothing). The continuum
flux density of 4C37.11 is 1.43 Jy.

The observed \HI\ absorption profile is shown in Fig.\ 1.  The extremely broad
profile is detected in both observations. The \HI\ absorption profile covers
about 1600 \kms\ (full width zero intensity - FWZI) and is dominated by two peaks: one centred on
$\sim$16000 \kms\ and a deeper, broader one centred on $\sim$17100 \kms.
Fainter absorption is also detected between these two peaks. Despite the
relatively similar bandwidth used, our \HI\ detection is much broader than
what found by Maness et al.\ (2004).  While the deep \HI\ absorption centered
around $\sim$17100 \kms\ is detected in their VLBI observations, the remaining
absorption does not appear in the profiles presented by them, although no
integrated profile is presented by Maness et al.\ (2004), so a direct
comparison with our data cannot be done. This means that for the deeper absorption we
know that it is located against the southern radio-lobe/hot-spot and extending
to about 20 pc, while the rest of the absorption may have been resolved out by
the high resolution o     f the VLBI data.  Part of the \HI\ absorption could be
against the more extended continuum emission (on scales $\sim$500 pc) that
appears in the 0.3 GHz VLBI image (Rodriguez et al.\ 2006). Another
possibility is that part of the absorption is against the kpc-scale radio
lobes described by Maness et al.\ (2000). However, as we will see in the next
Section, our optical spectra indicate that the \HI\ absorption occurs closer
to the AGN. As discussed for the
 case of Cygnus~A by Conway \& Blanco (1995),
velocity widths of the \HI\
 absorption broader than $\sim$150 \kms\ suggest
that the absorption is unlikely
 due to gas located at large distance from the
nucleus (i.e.\ considering the
 limited spatial scale of the continuum
background, it would imply  an overly large
 velocity gradient for gas associated
with a large-scale dust-lane). This
 argument appears to be even more valid in
the case of the broad absorption
 detected in 4C37.11.

 At the low resolution of our WSRT observations, 4C37.11 is unresolved and,
therefore, the low optical depth that we derive for the absorption centred on
$\sim$17100 \kms\ (peak $\tau \sim 0.005$) is a lower limit because the VLBA
data of Maness et al.\ show that the covering factor is less than 1. Indeed, 
Maness et al.\ (2004) derive an higher value for the peak optical depth of
this component of $\tau \sim 0.0179$ (and a column density of $1.8 \times
10^{20}$ cm$^{-2}$ for $T_{\rm spin}=100$K).

\label{sec:optical}
\section{The broad optical lines}

In order to compare  the kinematics of the  neutral and ionised gas, we have
performed new optical observations.
An optical spectrum of 4C37.11 was taken
with the William Herschel Telescope (WHT) on 13 January 2004 using the ISIS
long-slit spectrograph with the 6100 \AA\ dichroic, the R300B and R316R gratings
in the blue and red arm respectively and the GG495 blocking filter in the red
arm to cut out second-order blue light. This resulted in a wavelength coverage
from about 3500 to 8400 \AA. A slit with width of 1.3 arcsec has been used,
aligned along the parallactic angle of the observations (PA 120$^{\circ}$),
i.e.\  almost perpendicular to the PA of the radio source on VLBA-scales. The
observations were done at an airmass of $1.04$-$1.08$ and the seeing was
approximately 1.9 arcsec. The total integration time was 1800 sec for both the
blue and red arm. Here we present results of the red arm (6000 - 8400 \AA),
because  \OI, \HaNII\ and \SII\ are the only identifiable emission lines in the 
spectrum (see also Stickel et al 1993). We used the Image Reduction and Analysis Facility (IRAF) for a
standard reduction of the data. 
The accuracy of the  wavelength calibration is
0.3 \AA. 
The spectral resolution derived from  night sky-lines  is 4.2 \AA.

\begin{figure}
\centering
\includegraphics[width=0.40\textwidth]{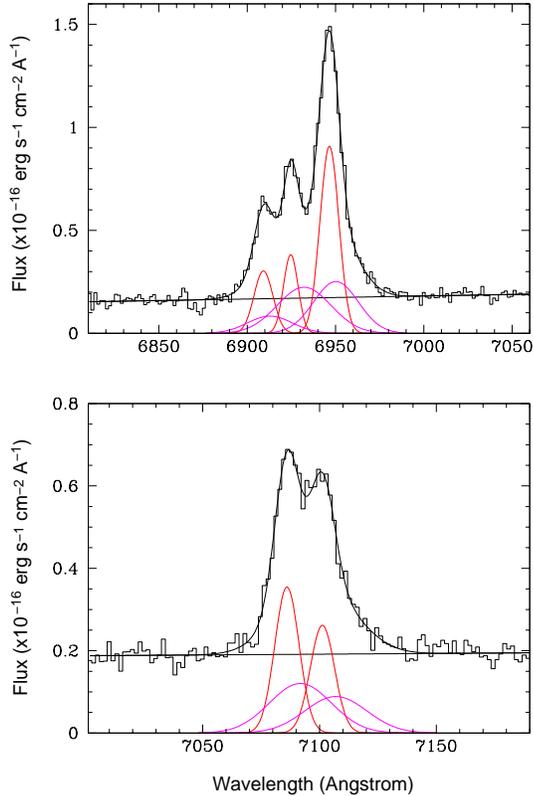}
\caption{Two-component fit to the \HaNII\ triplet (top) and \SII\ doublet (bottom). Details of the fit are given in Table \ref{tab:kinematics}}
\label{fig:HaNII}
\end{figure}

In order to analyse the kinematics of the emission-line gas, we fitted the
profile of the \HaNII\ emission-lines. This was done by using the Starlink
software to create 1D spectra and fit Gaussian profiles to the emission-lines.
Figure \ref{fig:HaNII} shows the \HaNII\ profile across an aperture of 1.6
arcsec in the central part of the host galaxy. Following atomic physics, the
Gaussian fit to the \HaNII\ triplet was done by constraining the width of the
\NII\ doublet components to be the same and their intensity ratios to be 3.0
while leaving all other parameters free. A good fit  was only obtained by using
two Gaussian components (see Fig.\ 2 - top).
The second component is much
broader and is slightly redshifted with respect to the first, narrower one, in
agreement with Rodriguez et al.\ (2006). 
 This result is also confirmed by 
fitting  the \SII\ doublet using a similar procedure (see Fig.\ 2 -
bottom).

The \HaNII\ triplet is spatially extended  out to a radius of at least
4 arcsec ($\sim$4 kpc), but the signal outside the central aperture is too weak
to accurately fit a 2-component model. No significant evidence for rotation is
present in the data (although we note that our spectra were not chosen to be
aligned with any kinematical  axis).

The systemic velocity derived from the
 average central velocity of the narrow
component of the \HaNII\ lines in the
 central region is $v_{\rm sys} = 16502
\pm 24$ \kms\ or $z = 0.055046 \pm
 0.000078$ (using the barycentric optical
velocity definition). This redshift is
 in agreement with the estimate by Xu
et al.\ (1994) and Rodriguez et al.\
 (2006).

Figure \ref{fig:HI} shows the
width and central velocity of the narrow and broad emission line components in
relation to the \HI\ absorption profile. Interestingly, the broad component
has comparable width to the absorption profile. Thus, as in many other 
radio galaxies similar to 4C37.11 (i.e.\ young, compact radio sources), the ionised
and neutral gas phases appear to have similar kinematical
characteristics. This suggests that the absorbing \HI\ gas is located in the
central (i.e.\ inner kpc) region of the galaxy where the optical spectrum was taken.

\begin{table*}
\caption{Ionised gas kinematics}
\centering
\label{tab:kinematics}
\begin{tabular}{lccccc}
\hline\hline
Spectral line & \multicolumn{2}{c}{Narrow component} & \multicolumn{2}{c}{Broad component} &   $\Delta v$ \\
        & v$_{\rm center}$ & v$_{\rm width}$ & v$_{\rm center}$ & v$_{\rm width}$ & \\
\hline
\NII\     & $16505 \pm 15$ & $544 \pm 26$ & $16674 \pm 83$  & $1353 \pm 126$ & $169 \pm 84$  \\  
H$\alpha$ & $16500 \pm 18$ & $398 \pm 36$ & $16845 \pm 303$ & $1634 \pm 779$ & $345 \pm 304$ \\  
\SII\     & $16469 \pm 18$ & $513 \pm 43$ & $16725 \pm 88$  & $1384 \pm 253$ & $256 \pm 90$  \\  
\hline
\end{tabular} 
\flushleft
{Notes -- All units are \kms\  and values have been corrected for instrumental resolution. $\Delta v$ is the difference between the central velocity of the narrow and the broad component. Results for \SII\ need to be taken with caution (because, due to the underlying stellar continuum, the intensity ratio of the broad component was constrained to be the same as that of the narrow component) and are merely included to show the reliability of our model for fitting \HaNII.}
\end{table*} 

\section{Discussion: gas kinematics and binary black-holes}
\label{sec:results}

\subsection{The unusual nature of the gas kinematics in 4C37.11}
\label{sec:discussHI}

The intriguing result presented above is the detection, in a radio source
possibly hosting one of the rare  examples of a binary black-hole, of  very
broad \HI\ absorption whose main feature is two absorption components separated
in velocity by $\sim$1100\kms. 
In order to be able to interpret this absorption, we first  discuss how the properties of this
absorption compare with what is seen in other radio sources.

Broad \HI\ absorption profiles (FWZI
$>$1000 \kms) have been detected in a handful of young or re-started powerful
radio sources  (see e.g.\ Morganti et al.\ 2005 and references therein). In this
respect, it is not completely surprising to find such  broad absorption in
4C37.11, a CSO which are typically considered to be  young radio sources where
the radio jet could still be embedded in a relatively dense medium, perhaps left
over from a merger event.  The FWZI width of the absorption profile in
4C37.11 is as large as that of the broadest profiles observed by Morganti et
al.\ (2005). However, the feature that makes 4C37.11 unusual is that the
absorption profile is centred on the systemic velocity. The 4 radio galaxies
observed by Morganti et al.\  that have HI profile widths in excess of 1000
km/s, all have profiles that are quite asymmetric and blueshifted relative to
the systemic velocity of the host galaxy. This difference means that the
absorption in 4C37.11 may not be connected to gas outflows, or if it is, some
key difference exists, e.g., different geometry.

On the other hand, symmetric \HI\ absorption profiles  centred on the systemic
velocity of the host galaxy have been found in many radio sources (e.g., Conway
\& Blanco 1995; Morganti et al.\ 2001, Vermeulen et al.\ 2003). However, the
typical width of these absorption profiles is of the order of $200$ \kms, much
narrower than what found in 4C37.11. Such relatively narrow symmetric absorption
is often thought to be caused by circumnuclear disks on the hundred-pc scale in
which the rotational velocity reflects the galactic potential.  The fact that
the absorption in 4C37.11 is much broader most likely  means that here it is not
caused by such a large structure.

It is interesting to note that the ionised gas also has a broad component 
centred close to the velocity of the narrow component, i.e.\ on the systemic
velocity. This is also different from other compact (steep spectrum) radio
sources (Holt et al.\ 2008 and refs therein).  In these young radio sources, the
broad components of the ionised gas are mainly  found to be blueshifted and
associated with outflows connected with the first phases of the nuclear
activity.

\subsection{Effects of binary BH on the gas kinematics?}

The central question is, of course, what causes the broad, two-component
absorption in 4C37.11.  Maness et al.\ (2004) argue that the \HI\ absorption
detected by them (i.e.\ only the redshifted component) is gas expelled by the
receding radio jet. Considering that the gas producing the \HI\ absorption
has to be located in front of the continuum, this can only happen if the
outflowing gas is accelerated and dragged by the jet cocoon (so that it
appears in front of the jet). Because now the \HI\ absorption is actually
found to be symmetric around the systemic velocity, this would then suggest
that also on the side of the main jet an outflow of similar amplitude is present. 
The fact that the source is classified as a compact {\sl symmetric} object could indeed suggest that the jets are oriented close to the plane of the sky, although this would imply very high flow velocities to explain the large observed (radial) component.  Moreover, Rodriguez et al.\ argue that the jets are oriented $<66^{\circ}$ from the line-of-sight, not quite in the plane of the sky. A further complication is that the location of the blueshifted absorption is not known and may not occur against the main jet. Nevertheless, given all uncertainties, a model explaining the two absorption components as approaching and receding outflows is probably not inconceivable. 
 
The discussion in Sec.\ \ref{sec:discussHI} shows that the properties of
4C37.11 are quite unusual, and perhaps this means that the absorption is
caused by a different mechanism. Maness et al.\ (2004) and Rodriguez et al.\
(2006) inferred from their observations that the most likely explanation for
the radio morphology of 4C37.11 is the presence of a binary system with two
 radio-loud, flat spectrum AGN separated by 7.3 pc.  The unique
characteristics of the very broad, two-component \HI\ absorption are perhaps a
further indication that extreme conditions are indeed present in this
object. It is conceivable that the absorbing gas is part of a structure in
rapid rotation as a result of the effect of two black holes orbiting. If so,
one can use the \HI\ velocity to estimate the mass of the two black holes
combined. In this case, we would have to assume that the blueshifted part of
the absorption is located against the (northern) lobe of 4C37.11 at $\sim$20
pc from the nucleus, i.e.\ symmetrical to the redshifted part detected in VLBI
by Manness et al.\ (2004). The velocity separation between the two peaks of
the \HI\ suggest that the orbital velocity is about $\sim$500 \kms. Using
these values, we find an estimated black holes mass of $\sim$$10^9$ $M_\odot$.
 
This mass compares very well with the mass of the central compact object
derived from the luminosity of the host galaxy, for example using the
relations presented in Ferrarese \& Ford (2005) and by Marconi \& Hunt (2003).
The observed $K$-band magnitude (2MASS) implies a black hole mass of just over
$10^9$ M$_{\odot}$. This is much higher than estimated in a similar way by
Rodriguez et al.\ (2006) using a $V$-band magnitude. However, the galactic
latitude of 4C37.11 is only $10.5^\circ$ and the correction for galactic
extinction is very uncertain. For example, the galactic $V$-band extinction
from Schlegel et al.\ (1998) is almost 3 magnitudes larger than that used by
Rodriguez et al.. If this higher extinction is used, a very similar value for
the combined BH mass is derived compared to the one found from the $K$
magnitude. It appears, therefore, that if the broad absorption is interpreted
as the kinematic signature of a binary black hole, it leads to a consistent
mass estimate. The weakness in the argument is that the location of the
blueshifted component in not known and this makes a detailed interpretation of
the data uncertain. It may indicate that the blueshifted part occurs on the
scale of many tens of pc (i.e.\ resolved out by the observations of Maness et
al.\ 2004). Even in this case the kinematics of the gas could still be
affected by the presence of the binary BH, but the kinematics would be more
difficult to interpret and deriving the BH mass would not be straightforward.

\section{Conclusions}

 We have presented the detection of surprisingly broad \HI\ absorption,
mainly consisting of two components separated in velocity by $\sim$1100 \kms,
against the compact symmetric object 4C37.11. Interestingly, this radio source
hosts a candidate supermassive binary black-hole system. The characteristics
of the \HI\ absorption are unusual compared to \HI\ absorption seen in other
radio sources. The presence of a fast outflow to explain the broad \HI\ absorption cannot be ruled out, but an alternative interpretation could be that it 
may indicate the existence of a binary black hole in 4C37.11. If interpreted
as being the kinematic signature of a binary black hole, it gives an
estimate of the mass of the combined black holes consistent with what derived from the luminosity of the spheroid. However, the fact that it is not
known where exactly the blueshifted absorption is occurring prevents an unambiguous interpretation. It is clear that further
high-resolution \HI\ observations of 4C37.11 are needed to determine what is
really occuring in the centre of 4C37.11.  Our data do show that the nucleus
of 4C37.11 is gas rich and, if 4C37.11 indeed harbours a binary black hole, it
underlines the importance of gas in the formation of such systems.

\begin{acknowledgements}

The WSRT is operated by the Netherlands Institute for Radio Astronomy
(ASTRON) with the support from the Netherlands Foundation for Scientific
Research (NWO). This research has made use of the NASA Extragalactic Database
(NED), whose contributions to this paper are gratefully acknowledged.  This publication makes use of data products from the Two Micron All Sky Survey, which is a joint project of the University of Massachusetts and the Infrared Processing and Analysis Center/California Institute of Technology, funded by the National Aeronautics and Space Administration and the National Science Foundation.

\end{acknowledgements}

{}

\end{document}